\documentclass[aps,showpacs,amsmath,amssymb]{revtex4}

\usepackage{latexsym}
\usepackage{graphicx}
\usepackage{dcolumn}
\usepackage{bm}
\usepackage{hyperref}

\newcommand{\beq}{\begin{equation}}
\newcommand{\eeq}{\end{equation}}
\newcommand{\beqa}{\begin{eqnarray}}
\newcommand{\eeqa}{\end{eqnarray}}
\newcommand{\beqar}{\begin{eqnarray*}}
\newcommand{\eeqar}{\end{eqnarray*}}
\newcommand{\bra}[1]{\mbox{$\left\langle{#1}\right|$}}
\newcommand{\ket}[1]{\mbox{$\left|{#1}\right\rangle$}}
\newcommand{\diracsp}[2]{\mbox{$\langle{#1}|{#2}\rangle$}}

\def\I{{\rm i}}
\def\d{{\rm d}}
\def\e{{\rm e}}

\newcounter{saveeqn}

\begin{document}

\title{Quantum mechanics in general quantum systems (IV): \\ Green operator and path integral}
\author{An Min Wang}\email{anmwang@ustc.edu.cn}
\affiliation{Quantum Theory Group, Department of Modern Physics,
University of Science and Technology of China, Hefei, 230026,
P.R.China}

\begin{abstract}

We first rewrite the perturbation expansion of the time evolution
operator [An Min Wang, quant-ph/0611216] in a form as concise as
possible. Then we derive out the perturbation expansion of the
time-dependent complete Green operator and prove that it is just the
Fourier transformation of the Dyson equation. Moreover, we obtain
the perturbation expansion of the complete transition amplitude in
the Feynman path integral formulism, and give an integral expression
that relates the complete transition amplitude with the unperturbed
transition amplitude. Further applications of these results can be
expected and will be investigated in the near future.

\end{abstract}


\maketitle

\section{Introduction}\label{sec1}

The study on quantum mechanics in general quantum systems is still
progressing. Recently, we have obtained the exact solution
\cite{My1}, perturbation theory \cite{My2} and open systems dynamics
\cite{My3} in general quantum systems. However, a whole theoretical
formulism requires us to investigate more problems and obtain more
conclusions. We continue our endeavor in this paper.

As is well-known, it is convenient and general to study and express
quantum mechanics in the operator form as possible using Dirac
symbol system \cite{diracpqm}. This requires us to rewrite our
results in such a concise form. Actually, it is helpful to study on
the formal theory of quantum mechanics in general quantum systems.
This is the first task and the starting point in this paper.

After the exact solution \cite{My1} and perturbation theory
\cite{My2} of quantum mechanics in general quantum systems, we
should consider the Green operator (function) \cite{SakuraiQM} and
the Feynman transition amplitude (propagator) of path integral
\cite{FeynmanPI} because many physical conclusions can be expressed
and calculated by them. Actually, the Green function approach
provides a unified systematic method for calculating various
quantities of physical interest. The path integral formulism builds
a formal frame for more complicated quantum theory, for example,
quantum field theory and quantum statistics, and gives a relatively
easy road to expressions for Green¡¯s functions, which are closely
related to amplitudes for physical processes such as scattering and
decays of particles.

Whatever Green operator or path integral, their calculations are not
simple in general when the concerned systems are complicated. Hence,
from the view of perturbation theory, we would like to obtain their
perturbation expansion for general quantum systems in order to avoid
the difficulty and carry out related calculations. At present, this
is a known effective way to deal with quantum mechanics in general
and complicated quantum systems. When we focus our attention on the
time evolution of the systems, it will be very useful to extend
Dyson equation \cite{DysonEQ,Joachain} about the time-independent
(stationary) Green operator (function) to one about the
time-dependent (dynamical) Green operator (function), that is, to
derive out an explicit perturbation expansion of the time-dependent
complete Green operator (function). Here, the time-independent Green
function means that it is a solution to the stationary Schr\"odinger
equation with the point resource, while the time-dependent Green
function means that it is a solution to the dynamical Schr\"odinger
equation with the point resource. Moreover, it is very interesting
to find the explicit perturbation expansion of the Feynman
transition amplitude (propagator) with the path integral form in
general quantum systems. Furthermore, we expect to relate the
complete transition amplitude to the unperturbed transition
amplitude. This must be able to provide a way to calculate the path
integral expression of the transition amplitude even the concerned
systems are complicated. In addition, the two perturbation
expansions include all order approximations of perturbation and have
a known general term, they should be able to play an important role
in the formal study of quantum theory.

This paper is organized as the following: in this section, we give
an introduction; in Sec. \ref{sec2}, we rewrite the perturbation
expansion of the time evolution operator \cite{My1} in the form as
concise as possible; in Sec. \ref{sec3}, we derive out the
perturbation expansion of the time-dependent complete Green operator
and prove that it is just a Fourier transformation of the Dyson
equation; in Sec. \ref{sec4}, we obtain the perturbation expansion
of the complete transition amplitude in the Feynman path integral
formulism, and give an integral expression that relates the complete
transition amplitude with the unperturbed transition amplitude; in
Sec. \ref{sec5}, we give some discussions and summarize our
conclusions.

\section{Perturbation expansion of the time evolution operator}\label{sec2}

In our recent paper \cite{My1}, we obtain the general and explicit
expression of the time evolution operator (for simplicity, set
$\hbar=1$) \beqa\label{eteo} \e^{-\I H t}&=&\sum_{l=0}^\infty
\mathcal{A}_l(t)=\sum_{l=0}^\infty\sum_{\gamma,\gamma^\prime}
A_l^{\gamma\gamma^\prime}(t)\ket{{\Phi}^{\gamma}}\bra{{\Phi}^{\gamma^\prime}}\eeqa
where the Hamiltonian $H$ has been written as the summation of the
unperturbed $H_0$ and the perturbing Hamiltonian $H_1$. While
$A_l^{\gamma\gamma^\prime}(t)$ are defined by
\beqa\label{Aldefinition} A_0^{\gamma\gamma^\prime}(t)&=&\e^{-\I
E_{\gamma}
t}\delta_{\gamma\gamma^\prime}\\
A_l^{\gamma\gamma^\prime}(t)&=&\sum_{\gamma_1,\cdots,\gamma_{l+1}}\left[
\sum_{i=1}^{l+1}(-1)^{i-1}\frac{\e^{-\I E_{\gamma_i}
t}}{d_i(E[\gamma,l])}\right]\left[
\prod_{j=1}^{l}H_1^{\gamma_j\gamma_{j+1}}\right]
\delta_{\gamma_1\gamma}\delta_{\gamma_{l+1}\gamma^\prime}.\eeqa
where $H_1^{\gamma_i\gamma_{i+1}}
=\bra{\Phi^{\gamma_{i}}}H_1\ket{\Phi^{\gamma_{i+1}}}$ are called the
perturbing Hamiltonian matrix elements, $\ket{\Phi^\gamma}$ is the
eigenvector of $H_0$ and $E_\gamma$ is the corresponding eigenvalue,
that is \beq\label{h0eeq} H_0\ket{\Phi^\gamma}=E_\gamma
\ket{\Phi^\gamma}, \eeq and the denominators $d_i(E[\gamma,l])$ in
$A_l^{\gamma\gamma^\prime}(t)$ reads \beqa
d_1(E[\gamma,l])&=&\prod_{i=1}^{l}\left(E_{\gamma_{1}}
-E_{\gamma_{i+1}}\right),\\
 d_i(E[\gamma,l])&=&
\prod_{j=1}^{i-1}\left(E_{\gamma_{j}}
-E_{\gamma_{i}}\right)\!\!\!\prod_{k=i+1}^{l+1}\left(E_{\gamma_{i}}
-E_{\gamma_{k}}\right),\\[-3pt] d_{l+1}(E[\gamma,l])
&=&\prod_{i=1}^{l}\left(E_{\gamma_{i}}-E_{\gamma_{l+1}}\right),\eeqa
where $2\leq i \leq l$. The expression of the time evolution
operator (\ref{eteo}) is both a $c$-number function and a power
series of the perturbation. Therefore, it is useful and effective
for the calculation of perturbation theory. However, it is not
convenient for the study on the formal theory of quantum theory. In
order to overcome this shortcoming, we, using the Dirac symbol
formulism, first rewrite the $\mathcal{A}$ in the following form
\beq \label{Acform}
\mathcal{A}_l(t)=\sum_{i=1}^{l+1}\sum_{\gamma_i}\left(G_{0E_{\gamma_i}}H_1\right)^{i-1}\e^{-\I
H_0 t}\ket{\Phi^{\gamma_i}}\bra{\Phi^{\gamma_i}}\left(H_1
G_{0E_{\gamma_i}}\right)^{l+1-i}\eeq where $G_{0E_{\gamma_i}}$ is
the unperturbed Green operator defined by \beq
G_{0E_{\gamma_i}}=\frac{1}{E_{\gamma_i}-H_0}\eeq or more strictly,
it is the restarted or advanced unperturbed Green operator defined
by \beq
G_{0E_{\gamma_i}}^{(\pm)}=\frac{1}{E_{\gamma_i}-H_0\pm\I\epsilon}\eeq
where $\epsilon\rightarrow 0^+$ \cite{Joachain} and $E_{\gamma_i}$
take over all of eigenvalues of $H_0$. Note that we cannot write the
expression of the time evolution operator in a full operator form
independent of the representation because its perturbation expansion
depends on the spectrum of the unperturbed Hamiltonian. Substituting
(\ref{Acform}) into (\ref{eteo}), we have the concise form of the
time evolution operator  \beqa\label{ceteo} \e^{-\I H
t}=\sum_{l=0}^\infty\sum_{i=1}^{l+1}\sum_{\gamma_i}\left(G_{0E_{\gamma_i}}^{(\pm)}H_1\right)^{i-1}\e^{-\I
H_0 t}\ket{\Phi^{\gamma_i}}\bra{\Phi^{\gamma_i}}\left(H_1
G_{0E_{\gamma_i}}^{(\pm)}\right)^{l+1-i}\eeqa

\section{Perturbation expansion of the time-dependent complete Green operator}\label{sec3}

In quantum mechanics, the Green operator (or function) plays a very
important role, for example, in the perturbation theory. It is a
kernel of the integral form of the Schr\"odinger equation, in
special, the Lippmann-Schwinger equation \cite{LSE}. It is clear
that we have \beq
\left[E_{\gamma}-H_0\right]G_{0E_{\gamma}}^{(\pm)}=1\eeq Or in the
representation of localized coordinate bases, the corresponding
Green function
$G_{0E_{\gamma}}^{(\pm)}(\bm{x},\bm{x}^\prime)=\bra{\bm{x}}G_{0E_{\gamma}}^{(\pm)}\ket{\bm{x}^\prime}$
obeys the stationary Schr\"oding equation with the point resource,
i.e \beq \left[E_{\gamma}-H_0(\bm{x},-\I
\bm{\nabla}_{\bm{x}})\right]G_{0E_{\gamma}}^{(\pm)}(\bm{x},\bm{x}^\prime)=\delta(\bm{x}-\bm{x}^\prime)\eeq
Here, we called it as the time-independent (stationary) unperturbed
Green function in order to distinguish the time-dependent
(dynamical) unperturbed Green function that obeys the following
dynamical Schr\"oding equation with the point resource: \beq
\left[\I\frac{\partial}{\partial t}-H_0(\bm{x},-\I
\bm{\nabla}_{\bm{x}})\right]G_{0}^{(\pm)}(\bm{x},\bm{x}^\prime;t,t^\prime)
=\delta(t-t^\prime)\delta(\bm{x}-\bm{x}^\prime)\eeq Similarly, the
time-independent complete Green function and the time-dependent
complete Green function satisfy, respectively \beq
\left[E_{\gamma}-H(\bm{x},-\I
\bm{\nabla}_{\bm{x}})\right]G_{E_{\gamma}}^{(\pm)}(\bm{x},\bm{x}^\prime)=\delta(\bm{x}-\bm{x}^\prime)\eeq
\beq \left[\I\frac{\partial}{\partial t}-H(\bm{x},-\I
\bm{\nabla}_{\bm{x}})\right]G^{(\pm)}(\bm{x},\bm{x}^\prime;t,t^\prime)
=\delta(t-t^\prime)\delta(\bm{x}-\bm{x}^\prime)\eeq It is easy to
obtain the time-independent complete Green operator \beq
G_{E_{\gamma}}^{(\pm)}=\frac{1}{E_{\gamma}-H\pm\I\epsilon}\eeq
Moreover, the time-independent complete Green operator (or function)
can be given by the Dyson equation \beq\label{Dysoneq}
G_{E_{\gamma}}^{(\pm)}=G_{0E_{\gamma}}^{(\pm)}+G_{0E_{\gamma}}^{(\pm)}H_1
G_{E_{\gamma}}^{(\pm)}=G_{0E_{\gamma}}^{(\pm)}\sum_{l=0}^\infty
\left(H_1G_{0E_{\gamma}}^{(\pm)}\right)^l\eeq It is actually the
perturbation expansion of the time-independent complete Green
operator. Just well-known, the time-dependent Green operator is the
Fourier transformation of the time-independent Green operator, that
is \beqa\label{cg0d}
G_0^{(\pm)}(t,t^\prime)&=&\frac{1}{2\pi}\int_{-\infty}^\infty \d
E_\gamma
G_{0E_\gamma}^{(\pm)}\e^{-\I E_\gamma(t-t^\prime)}\\
\label{cgd}
G^{(\pm)}(t,t^\prime)&=&\frac{1}{2\pi}\int_{-\infty}^\infty \d
E_\gamma G_{E_\gamma}^{(\pm)}\e^{-\I E_\gamma(t-t^\prime)} \eeqa
Obviously, it is difficult to calculate this integral when we
directly substitute the Dyson equation (\ref{Dysoneq}) into the
above relation. In other words, ones have not yet clearly known its
explicit form, that is, a perturbation expansion with the factors of
primary functions of time so far. However, using the concise form of
the time evolution operator (\ref{ceteo}), we can derive out it.

Actually, this derivation is easy from the relation between the
time-dependent complete Green operator and the time evolution
operator \beqa G^{(+)}(t,t^\prime)&=&-\I\theta(t-t^\prime)\e^{-\I
H(t-t^\prime)}\\
G^{(-)}(t,t^\prime)&=&\I\theta(t^\prime-t)\e^{-\I
H(t-t^\prime)}\eeqa This immediately yields \beqa \label{peofgo}
G^{(+)}(t,t^\prime)&=&-\I\theta(t-t^\prime)\sum_{l=0}^\infty\sum_{i=1}^{l+1}
\sum_{\gamma_i}\left(G_{0E_{\gamma_i}}^{(+)}H_1\right)^{i-1}\e^{-\I
H_0 (t-t^\prime)}\ket{\Phi^{\gamma_i}}\bra{\Phi^{\gamma_i}}\left(H_1
G_{0E_{\gamma_i}}^{(+)}\right)^{l+1-i}\\
G^{(-)}(t,t^\prime)&=&\I\theta(t^\prime-t)\sum_{l=0}^\infty\sum_{i=1}^{l+1}
\sum_{\gamma_i}\left(G_{0E_{\gamma_i}}^{(-)}H_1\right)^{i-1}\e^{-\I
H_0 (t-t^\prime)}\ket{\Phi^{\gamma_i}}\bra{\Phi^{\gamma_i}}\left(H_1
G_{0E_{\gamma_i}}^{(-)}\right)^{l+1-i}\eeqa

Now, let us prove the above perturbation expansion of the
time-dependent complete Green operator be a Fourier transformation
of the Dyson equation (\ref{Dysoneq}). In other words, the inversion
transformation of Eq. (\ref{peofgo}) is just the Dyson equation
(\ref{Dysoneq}).

From the relations
\beqa -\I\int_{-\infty}^\infty \d t
\;\theta(t-t^\prime)\e^{-\I H_0 (t-t^\prime)} \e^{\I E_{\gamma}
(t-t^\prime)}&=&G_{0E_{\gamma}}^{(+)}\\
\I\int_{-\infty}^\infty \d t \;\theta(t^\prime-t)\e^{-\I H_0
(t-t^\prime)} \e^{\I E_{\gamma}
(t-t^\prime)}&=&G_{0E_{\gamma}}^{(-)} \eeqa it follows that \beqa
G_{E_{\gamma}}^{(\pm)}&=& \int_{-\infty}^\infty \d t\;
G^{(\pm)}(t,t^\prime)\e^{\I E_{\gamma} (t-t^\prime)}\nonumber\\
&=&\sum_{l=0}^\infty\sum_{i=1}^{l+1}
\sum_{\gamma_i}\left(G_{0E_{\gamma_i}}^{(\pm)}H_1\right)^{i-1}G_{0E_{\gamma}}^{(\pm)}
\ket{\Phi^{\gamma_i}}\bra{\Phi^{\gamma_i}}\left(H_1
G_{0E_{\gamma_i}}^{(\pm)}\right)^{l+1-i}\nonumber\\
&=&-\sum_{l=0}^\infty\sum_{\gamma,\gamma^\prime}
\sum_{\gamma_1,\cdots,\gamma_{l+1}}\left[
\sum_{i=1}^{l+1}(-1)^{i-1}\frac{1}{d_i^{(\pm)}(E[\gamma,l])(E_{\gamma_i}-E_\gamma\mp\I\epsilon)}\right]\left[
\prod_{j=1}^{l}H_1^{\gamma_j\gamma_{j+1}}\right]
\delta_{\gamma_1\gamma}\delta_{\gamma_{l+1}\gamma^\prime}
\ket{{\Phi}^{\gamma}}\bra{{\Phi}^{\gamma^\prime}}\eeqa where
$d_i^{(\pm)}(E[\gamma,l])$ is defined by adding $\pm\I\epsilon$ to
its every factor of the energy level difference.

Based on the appendix in our previous paper \cite{My1}, if we set
$E_\gamma=E_{\gamma_{l+2}}\mp 2\I\epsilon$, we have
$d_i(E[\gamma,l])(E_{\gamma_i}-E_{\gamma_{l+2}})=d_i(E[\gamma,l+1])$.
Again using our identity \cite{My1} \beq \label{myi}\sum_{i=1}^{l+1}
(-1)^{i-1} \frac{E_{\gamma_i}^K}{d_i(E[\gamma,l])}
=\left\{\begin{array}{c l}0 &\quad (\mbox{If $0\leq K<l$})\\[8pt] 1 &\quad (\mbox{If
$K=l$})\end{array}\right.. \eeq we obtain \beqa
G_{E_{\gamma}}^{(\pm)}
&=&G_{0E_{\gamma}}^{(\pm)}+\sum_{l=1}^\infty\sum_{\gamma,\gamma^\prime}
\sum_{\gamma_1,\cdots,\gamma_{l+1}}\left.
(-1)^{l+1}\frac{1}{d_{l+2}^{(\pm)}(E[\gamma,l+1])}\right|_{E_{\gamma_{l+2}}=E_{\gamma}\pm
2\I\epsilon} \left[ \prod_{j=1}^{l}H_1^{\gamma_j\gamma_{j+1}}\right]
\delta_{\gamma_1\gamma}\delta_{\gamma_{l+1}\gamma^\prime}
\ket{{\Phi}^{\gamma}}\bra{{\Phi}^{\gamma^\prime}}\\
&=& G_{0E_{\gamma}}^{(\pm)}\sum_{l=0}^\infty
\left(H_1G_{0E_{\gamma}}^{(\pm)}\right)^l=G_{0E_{\gamma}}^{(\pm)}+G_{0E_{\gamma}}^{(\pm)}H_1
G_{E_{\gamma}}^{(\pm)}\eeqa Therefore, we finish our proof. This
implies that the perturbation expansion (\ref{peofgo}) of the
time-dependent complete Green operator, as the Fourier
transformation of the Dyson equation (\ref{Dysoneq}), is obtained by
using the concise form of the time evolution operator (\ref{ceteo}).

\section{Perturbation expansion of the complete transition
amplitude}\label{sec4}

Path integral is a formulism that yields the quantum-mechanical
amplitudes in a global approach involving the properties of a system
at all times. From our perturbation expansion of the time evolution
operator, we can see that the time-dependent parts for every term in
the summation (\ref{ceteo}) are expressed in a factor $\exp(-\I H_0
t)$ . Hence, only to find the path integral expression of $\exp(-\I
H_0 t)$ can keep the features and advantages of the path integral
formula.

Here, for simplicity, we assume the space to be one-dimensional.
Thus, the unperturbed transition amplitude $K_0(x_b,t_b;x_a,t_a)$
and the complete transition amplitude $K(x_b,t_b;x_a,t_a)$
($t_b>t_a$) are defined, respectively, by\beqa
K_0(x_b,t_b;x_a,t_a)&=&\bra{x_b}\e^{-\I H_0 (t_b-t_a)}
\ket{x_a}\\
K(x_b,t_b;x_a,t_a)&=&\bra{x_b}\e^{-\I H (t_b-t_a)} \ket{x_a}\eeqa
From Eq. (\ref{ceteo}) it immediately follows that the perturbation
expansion of the complete transition amplitude \beq
K(x_b,t_b;x_a,t_a)=\sum_{l=0}^\infty\sum_{i=1}^{l+1}\sum_{\gamma_i}
\int\d y_b \d y_a
\bra{x_b}\left(G_{0E_{\gamma_i}}^{(\pm)}H_1\right)^{i-1}\ket{y_b}K_0(y_b,t_b;y_a,t_a)
\diracsp{y_a}{\Phi^{\gamma_i}}\bra{\Phi^{\gamma_i}}\left(H_1
G_{0E_{\gamma_i}}^{(\pm)}\right)^{l+1-i}\ket{x_a}\eeq In fact, this
is an integral expression that relates the complete transition
amplitude with the unperturbed transition amplitude, that is \beq
\label{ieofta} K(x_b,t_b;x_a,t_a)= \int\d y_b \d y_a
C(x_b,y_b;x_a,y_a)K_0(y_b,t_b;y_a,t_a)\eeq where \beq
C(x_b,y_b;x_a,y_a)=\sum_{l=0}^\infty\sum_{i=1}^{l+1}\sum_{\gamma_i}
\bra{x_b}\left(G_{0E_{\gamma_i}}^{(\pm)}H_1\right)^{i-1}\ket{y_b}
\diracsp{y_a}{\Phi^{\gamma_i}}\bra{\Phi^{\gamma_i}}\left(H_1
G_{0E_{\gamma_i}}^{(\pm)}\right)^{l+1-i}\ket{x_a}\eeq If the path
integral expression of $K_0(x_b,t_b;x_a,t_a)$ has been found as
follows \cite{FeynmanPI,Kleinert} \beq \label{K0pie}
K_0(x_b,t_b;x_a,t_a)=A \int_{x(t_a)=x_a}^{x(t_b)=x_b} \mathcal{D} x
\e^{\I\mathcal{S}_0[x]}\eeq where $A$ and $S_0[x]$ are known, while
\beq \int\mathcal{D}x=\lim_{N\rightarrow\infty}\prod_{n=1}^N \d
x_n\eeq Substituting Eq. (\ref{K0pie}) into Eq. (\ref{ieofta}), we
can rewrite the integral expression (\ref{ieofta}) as \beq
\label{eieofta} K(x_b,t_b;x_a,t_a)=A \int\d y_b \d y_a
C(x_b,y_b;x_a,y_a)\int_{y(t_a)=y_a}^{y(t_b)=y_b} \mathcal{D} y
\e^{\I\mathcal{S}_0[y]} \eeq This implies that, if the unperturbed
transition amplitude $K_0(x,t;x^\prime,t^\prime)$ is known, then the
complete transition amplitude can be obtained in principle even the
concerned system is complicated. However, for practical purposes,
the cut-off approximation is possible to be required, and finding
the above integral analytically might not be a simple task. Perhaps,
the numerical method has its playing role.

Obviously, the extension of the above conclusions to three
dimensional space is direct.

\section{Conclusion and Discussion}\label{sec5}

In this paper, we first rewrite the perturbation expansion of the
time evolution operator in a more concise form than Ref. \cite{My1}.
Since this expansion depends on the whole spectrum of the
unperturbed Hamiltonian, it cannot be written as a full operator
form independent of the representation. But, this concise form
(\ref{ceteo}) indeed has been approximatively expressed as the
operator form. In terms of it, we also can rewrite our exact
solution. Such a form can be used to the our future study on the
formal theory of quantum mechanics in general quantum systems.

Then we derive out the perturbation expansion of the time-dependent
complete Green operator. It is difficult to obtain directly by the
definitions (\ref{cg0d}) and (\ref{cgd}) and the Dyson equation.
Moreover, we prove that it is a Fourier transformation of Dyson
equation. It is worthy pointing out that the general term of this
perturbation expansion has been found and all order approximation of
perturbation is included. Just like the case that Dyson equation has
extensive applications, we think that our perturbation expansion of
the time-dependent complete Green operator will have corresponding
interesting applications in the quantum theory, in special, in the
problems about the transition probability and the formal scattering.

In the last, we obtain the perturbation expansion of the complete
transition amplitude in the Feynman path integral formulism, and
give an integral expression that relates the complete transition
amplitude with the unperturbed transition amplitude. In fact, this
expansion will not only be able to simplify the calculation of path
integral in some quantum systems but also can provide a kind of way,
at least the probability, to obtain the expression of path integral
for general quantum systems. Although the result is an infinite
series, it can be cut-off since this series is a power series of
perturbation. For practical calculation, we are sure that this
expansion can produce the significant physical conclusions, at least
via the numerical methods. In addition, it may be interesting to
study the relation between our perturbation expansion of the
transition amplitude and the variational perturbation expansion of
the transition amplitude \cite{Kleinert}. However, since our
perturbation expansion provides a general term for any order
approximation, it should have its obvious advantages. It is clear
that our integral expression that relates the complete transition
amplitude with the unperturbed transition amplitude is helpful to
formally studies on quantum theory since it includes all order
approximations of perturbation in a tidy and explicit form.

At present, we mainly focus on our attention to try to build the
theoretical formulism of quantum mechanics in general quantum
systems. Limited by our time and capability, we will delay to study
the concrete applications to the near future.


\section*{Acknowledgments}

We are grateful all the collaborators of our quantum theory group in
the Institute for Theoretical Physics of our university. This work
was funded by the National Fundamental Research Program of China
under No. 2001CB309310, and partially supported by the National
Natural Science Foundation of China under Grant No. 60573008.


\end{document}